\documentclass[aps,PRB,twocolumn,superscriptaddress,preprintnumbers]{revtex4-2}
\usepackage[T1]{fontenc}
\usepackage{times}
\usepackage{graphicx,color}
\usepackage{amsfonts,amsmath,amssymb,amsbsy}
\usepackage[colorlinks=true, linkcolor=blue, citecolor=blue, urlcolor=blue]{hyperref}
\usepackage{float}

\begin{document}

\title{Intrinsic nonlinear conductivity induced by quantum geometry in
altermagnets\\
and measurement of the in-plane N\'{e}el vector}
\author{Motohiko Ezawa}
\affiliation{Department of Applied Physics, The University of Tokyo, 7-3-1 Hongo, Tokyo
113-8656, Japan}

\begin{abstract}
The $z$-component of the N\'{e}el vector is measurable by the anomalous Hall
conductivity in altermagnets because time reversal symmetry is broken. On
the other hand, it is a nontrivial problem how to measure the in-plane
component of the N\'{e}el vector. We study the second-order nonlinear
conductivity of a system made of the $d$-wave altermagnet with the Rashba
interaction. It is shown that the quantum-metric induced nonlinear
conductivity and the nonlinear Drude conductivity are proportional to the
in-plane component of the N\'{e}el vector, and hence, the in-plane component
of the N\'{e}el vector is measurable. We obtain analytic formulas of the
quantum-metric induced nonlinear conductivity and the nonlinear Drude
conductivity both for the longitudinal and transverse conductivities. The
quantum-metric induced nonlinear conductivity diverges at the Dirac point,
while the nonlinear Drude conductivity is always finite. Hence, the
quantum-metric induced nonlinear conductivity is dominant at the Dirac point
irrespective of the relaxation time.
\end{abstract}

\maketitle

\textbf{Introduction:} Altermagnets attracts much attention in the context
of spintronics\cite{SmejRev,SmejX,SmejX2}. One of the reason is that the
spin current can be generated\cite{Naka,Gonza,NakaB,Bose} without using the
spin-orbit interaction due to the momentum dependent band structure\cite%
{AhnAlter,Hayami,SmejRev,SmejX,SmejX2}. Indeed, momentum dependent band
structures are observed by Angle-Resolved Photo-Emission Spectroscopy (ARPES)%
\cite{Krem,Lee,Fed,Osumi,Lin}. Although the net magnetization is zero in
altermagnets, the $z$-component of the N\'{e}el vector is measurable with
the aid of anomalous Hall effects due to the breaking of time-reversal
symmetry\cite{Fak,Tsch,Sato,Leiv}.

Recently, there are many works on the nonlinear conductivity\cite%
{Gao,HLiu,Michishita,Watanabe,CWang,HLiu2,Oiwa,AGao,NWang,KamalDas,Kaplan,Ohmic,Xiang}%
. The second-order nonlinear conductivity $\sigma ^{ab;c}$ is defined by $%
j^{c}=\sigma ^{ab;c}E^{a}E^{b}$, where $E^{a}$ is an applied electric field
along the $a$ direction and $j^{c}$ is the current along the $c$ direction.
Especially, the nonlinear conductivity $\sigma ^{ab;c}$ has three
contributions: They are the quantum-metric induced nonlinear conductivity%
\cite{Kaplan} $\sigma _{\text{Metric}}^{ab;c}$, the Berry curvature dipole
induced nonlinear conductivity\cite{Sodeman} $\sigma _{\text{Dipole}}^{ab;c}$%
, and the nonlinear Drude conductivity\cite{Ideue,NLDrude} $\sigma _{\text{%
NLDrude}}^{ab;c}$. Here, quantum metric is defined by quantum distance with
respect to the wave functions\cite{Provost,Ma,Resta}. $\sigma _{\text{Dipolec%
}}^{ab;c}$\ and $\sigma _{\text{NLDrude}}^{ab;c}$ are proportional to $\tau $
and $\tau ^{2}$, respectively, where $\tau $ is the electron relaxation
time. They are extrinsic conductivities. On the other hand, $\sigma _{\text{%
Metric}}^{ab;c}$ is independent of $\tau $, which is an intrinsic nonlinear
conductivity. It is theoretically shown that $\sigma _{\text{Metric}}^{ab;c}$
is nonzero for the tilted Dirac system\cite{KamalDas} and the Rashba system
under in-plane magnetic field\cite{Teresa}. There is an experimental
observation of $\sigma _{\text{Metric}}^{ab;c}$ in a Rashba system\cite{Sala}%
. Furthermore, it is pointed out\cite{YFang} that the leading term of the
nonlinear conductivity is the third-order in $d$-wave altermagnets, where
the direction of the N\'{e}el vector is assumed to be along the $z$\
direction.

In this paper, we study the second-order nonlinear conductivity in a system
made of the $d$-wave altermagnet with the Rashba interaction without
applying magnetic field. It is shown to be proportional to the in-plane
component of the N\'{e}el vector in the $d$-wave altermagnet, and hence it
is measurable by measuring the second-order nonlinear conductivity. We
obtain analytic formula of the quantum-metric induced nonlinear conductivity
and the nonlinear Drude conductivity both for the longitudinal and
transverse conductivities by using the first-order perturbation theory with
respect to the magnitude of the altermagnetization. The quantum-metric
induced nonlinear conductivity diverges at the Dirac point both for the
longitudinal and transverse conductivities, while the nonlinear Drude
conductivity is always finite. Hence, the quantum-metric induced nonlinear
conductivity is dominant at the Dirac point irrespective of the relaxation
time.

\textbf{Nonlinear conductivity:} The second-order nonlinear conductivity $%
\sigma ^{ab;c}$ is expanded in terms of the electron relaxation time $\tau $
as\cite{Kaplan}

\begin{equation}
\sigma ^{ab;c}=\sigma _{\text{Metric}}^{ab;c}+\sigma _{\text{Dipole}%
}^{ab;c}+\sigma _{\text{NLDrude}}^{ab;c},
\end{equation}%
where%
\begin{equation}
\sigma _{\text{Metric}}^{ab;c}\propto \tau ^{0},\quad \sigma _{\text{Dipole}%
}^{ab;c}\propto \tau ,\quad \sigma _{\text{NLDrude}}^{ab;c}\propto \tau ^{2}.
\end{equation}%
We explain each term.

First, only the term $\sigma _{\text{Metric}}^{ab;c}$ survives in the dirty
limit $\tau \rightarrow 0$, which is the intrinsic nonlinear conductivity.
It is the quantum-metric induced nonlinear conductivity given by 
\begin{equation}
\sigma _{\text{Metric}}^{ab;c}\left( \mu \right) =-\frac{e^{3}}{\hbar }%
\sum_{n}\int d^{2}{k}f_{n}\left( 2\frac{\partial G_{n}^{ab}}{\partial k_{c}}-%
\frac{1}{2}\left( \frac{\partial G_{n}^{bc}}{\partial k_{a}}+\frac{\partial
G_{n}^{ac}}{\partial k_{b}}\right) \right) ,
\end{equation}%
where $f_{n}=1/\left( \exp \left( E_{n}-\mu \right) +1\right) $ is the Fermi
distribution function for the band $n$, $\mu $ is the chemical potential,
and $G_{n}^{ab}$ is the band--energy normalized quantum metric or the Berry
connection polarizability. It is given by\cite%
{Gao,HLiu,CWang,HLiu2,KamalDas,Kaplan,Teresa}%
\begin{equation}
G_{n}^{ab}=2\text{Re}\sum_{m\neq n}\frac{A_{nm}^{a}\left( \mathbf{k}\right)
A_{mn}^{b}\left( \mathbf{k}\right) }{\varepsilon _{n}\left( \mathbf{k}%
\right) -\varepsilon _{m}\left( \mathbf{k}\right) },  \label{BMetric}
\end{equation}%
with $\varepsilon _{n}$ being the energy of the band $n$, and $A_{nm}^{a}$\
being the interband Berry connection%
\begin{equation}
A_{nm}^{a}\left( \mathbf{k}\right) =i\left\langle \psi _{n}\left( \mathbf{k}%
\right) \right\vert \partial _{k_{a}}\left\vert \psi _{m}\left( \mathbf{k}%
\right) \right\rangle .
\end{equation}%
In what follows, we focus on the generic two-band system described by the
Hamiltonian%
\begin{equation}
H\left( \mathbf{k}\right) =h_{0}\left( \mathbf{k}\right) \sigma
_{0}+\sum_{j=x,y,z}h_{j}\left( \mathbf{k}\right) \sigma _{j},
\end{equation}%
where $\sigma _{j}$ is the Pauli matrix. The energy spectrum is given by%
\begin{equation}
E_{\pm }\left( \mathbf{k}\right) =h_{0}\left( \mathbf{k}\right) \pm h\left( 
\mathbf{k}\right) ,
\end{equation}%
where $h\left( \mathbf{k}\right) \equiv \sqrt{\sum_{j=x,y,z}h_{j}^{2}\left( 
\mathbf{k}\right) }$, by taking $n=\pm $ in Eq.(\ref{BMetric}). Eq.(\ref%
{BMetric}) is rewritten as\cite{Teresa}%
\begin{equation}
G_{-}^{ab}\left( \mathbf{k}\right) =2\text{Re}\frac{A_{-+}^{a}\left( \mathbf{%
k}\right) A_{+-}^{b}\left( \mathbf{k}\right) }{\varepsilon _{-}\left( 
\mathbf{k}\right) -\varepsilon _{+}\left( \mathbf{k}\right) }=-\frac{%
g^{ab}\left( \mathbf{k}\right) }{2h\left( \mathbf{k}\right) }
\end{equation}%
in terms of the quantum metric $g^{ab}$. For the two-band system it is given
by\cite{Matsuura,Gers,Onishi,WChen2024,EzawaGeo}%
\begin{equation}
g^{ab}\left( \mathbf{k}\right) =\frac{1}{2}\sum_{j=x,y,z}\left( \partial
_{k_{a}}h_{j}\right) \left( \partial _{k_{b}}h_{j}\right) .
\end{equation}%
Since this does not include $h_{0}\left( \mathbf{k}\right) $, $%
G_{-}^{ab}\left( \mathbf{k}\right) $ is independent of $h_{0}\left( \mathbf{k%
}\right) $.

\begin{figure}[t]
\centerline{\includegraphics[width=0.48\textwidth]{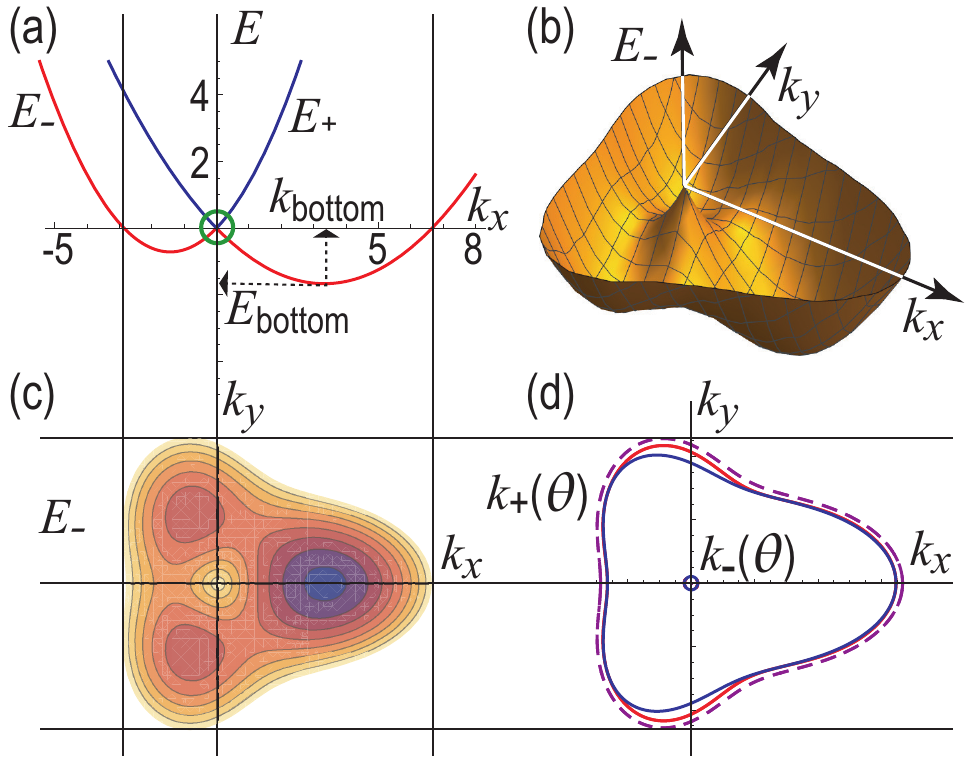}}
\caption{(a) Energy spectrum $E_{\pm }\left( k_{x},0\right) $ in units of $%
E_{0}=M\protect\lambda ^{2}/2\hbar ^{2}$. The horizontal axis is $k_{x}$ in
units of $k_{0}=M\protect\lambda /2\hbar ^{2}$. The green circle indicates
the Dirac point. (b) Bird's eye's view of the lower energy $E_{-}$ . (c)
Contour plot of the lower energy $E_{-}$ . (d) The tiny inner Fermi surface
is $k_{-}\left( \protect\theta \right) $ and the outer Fermi surfaces are $%
k_{+}\left( \protect\theta \right) $ at $\protect\mu =-0.2E_{0}$, where the
blue curve is an analytical solution (\protect\ref{kpm}) based on the
perturbation theory with respect to $J$, while the red curve is an analytic
solution without perturbation. Purple dashed curve indicates the Fermi
surface at $\protect\mu =0$. We have set $J=0.1E_{0}/k_{0}^{2}$.}
\label{FigBand}
\end{figure}

Second, $\sigma _{\text{Dipole}}^{ab;c}$ is the nonlinear transverse (Hall)
conductivity induced by the Berry curvature dipole\cite{Sodeman},%
\begin{equation}
\sigma _{\text{Dipole}}^{ab;c}\left( \mu \right) =-\frac{e^{3}\tau }{\hbar
^{2}}\sum_{n}\int d^{2}kf_{n}\left( \frac{\partial \Omega _{n}^{bc}}{%
\partial k_{a}}+\frac{\partial \Omega _{n}^{ac}}{\partial k_{b}}\right)
\end{equation}%
with the Berry curvature%
\begin{equation}
\Omega _{n}^{ab}\equiv \partial _{a}A_{nn}^{b}\left( \mathbf{k}\right)
-\partial _{b}A_{nn}^{a}\left( \mathbf{k}\right) .
\end{equation}%
It is an extrinsic nonlinear conductivity, since it vanishes as $\tau
\rightarrow 0$.

Third, $\sigma _{\text{NLDrude}}^{\text{ab;c}}$ is the nonlinear Drude
conductivity\cite{NLDrude},%
\begin{equation}
\sigma _{\text{NLDrude}}^{\text{ab;c}}\left( \mu \right) =-\frac{e^{3}\tau
^{2}}{\hbar ^{3}}\sum_{n}\int d^{2}kf_{n}\frac{\partial ^{3}E_{n}}{\partial
k_{a}\partial k_{b}\partial k_{c}},
\end{equation}%
where $E_{n}$ is the energy of the band $n$. It is also an extrinsic
nonlinear conductivity.

\textbf{Model:} We consider a system made of the $d$-wave altermagnets with
the Rashba interaction, whose Hamiltonian is given by\cite%
{SmejRev,SmejX,SmejX2}%
\begin{equation}
H\left( \mathbf{k}\right) =\frac{\hbar ^{2}\left( k_{x}^{2}+k_{y}^{2}\right) 
}{2M}\sigma _{0}+\lambda \left( k_{x}\sigma _{y}-k_{y}\sigma _{x}\right)
+J\left( k_{x}^{2}-k_{y}^{2}\right) \mathbf{n}\cdot \boldsymbol{\sigma },
\end{equation}%
where $M$ is the effective mass, $\lambda $ is the magnitude of the Rashba
interaction, $J$ is the magnitude of the $d$-wave altermagnetization, and $%
\mathbf{n}$ is the N\'{e}el vector of the $d$-wave altermagnet. The Rashba
interaction is introduced by placing an altermagnet on the substrate. The
Rashba interaction breaks inversion symmetry, while the altermagnet term
breaks time-reversal symmetry. Hence, the system breaks both inversion
symmetry and time-reversal symmetry. Then, the nonlinear Drude conductivity
and the quantum-metric induced nonlinear conductivity may emerge\cite{Kaplan}%
. We assume $\left\vert J\right\vert <\hbar ^{2}/\left( 2M\right) $ so that
the parabolic dispersion is positive for large $k=|\mathbf{k}|$.

The nonlinear conductivity $\sigma ^{yy;x}$ is nonzero when the N\'{e}el
vector is along the $y$ direction $\mathbf{n}=\left( 0,1,0\right) $, as we
show later in Eqs.(\ref{yyx1}) and (\ref{yyx2}).\ We take $\mathbf{n}=\left(
0,1,0\right) $ in the following. In this case, the energy is given by%
\begin{equation}
E_{\pm }\left( \mathbf{k}\right) =\frac{\hbar ^{2}k^{2}}{2M}\pm \sqrt{\left(
\lambda k_{x}+J\left( k_{x}^{2}-k_{y}^{2}\right) \right) ^{2}+\lambda
^{2}k_{y}^{2}},  \label{Energy}
\end{equation}%
which is shown in Fig.\ref{FigBand}. The inversion symmetry with respect to $%
k_{x}$ is broken for $J\neq 0$.

The energy along $k_{y}=0$ is shown in Fig.\ref{FigBand}(a). Two energy
spectra $E_{\pm }$ touch at the Dirac point $k_{x}=k_{y}=0$ and $E=0$, which
does not shift in the presence of altermagnetization $J$.

\begin{figure}[t]
\centerline{\includegraphics[width=0.48\textwidth]{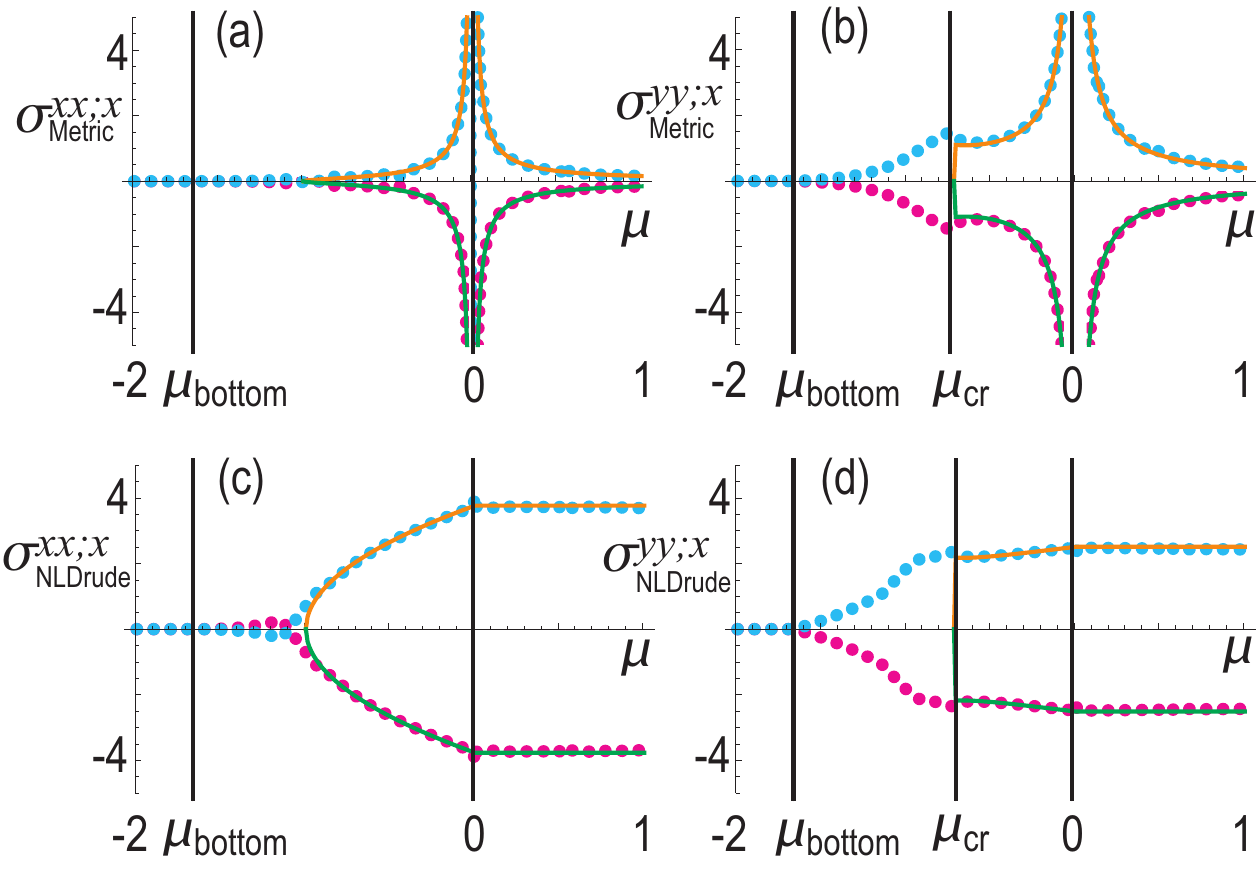}}
\caption{(a) Quantum-metric induced longitudinal conductivity $\protect%
\sigma _{\text{Metric}}^{xx;x}$ in units of $e^{3}/\left( \hbar
E_{0}k_{0}\right) $. (b) Quantum-metric induced transverse conductivity $%
\protect\sigma _{\text{Metric}}^{yy;x}$ in units of $e^{3}/\left( \hbar
E_{0}k_{0}\right) $. (c) Longitudinal nonlinear Drude conductivity $\protect%
\sigma _{\text{NLDrrude}}^{xx;x}$ in units of $e^{3}\protect\tau %
^{2}E_{0}/\left( \hbar ^{3}k_{0}\right) $. (d) Transverse nonlinear Drude
conductivity $\protect\sigma _{\text{NLDrrude}}^{yyx;}$ in units of $e^{3}%
\protect\tau ^{2}E_{0}/\left( \hbar ^{3}k_{0}\right) $. The horizontal axis
is the chemical potential $\protect\mu$ in units of $E_{0}$. Red dots
indicate numerically obtained results by setting $J=0.1E_{0}/k_{0}^{2}$,
while cyan dots indicate numerically obtained results by setting $%
J=-0.1E_{0}/k_{0}^{2}$. Green curves indicate analytically obtained results
by setting $J=0.1E_{0}/k_{0}^{2}$, while orange curves indicate analytically
obtained results by setting $J=-0.1E_{0}/k_{0}^{2}$. See the definition of $%
E_{0}$ and $k_{0}$ in the caption of Fig.1. We have set $\protect\mu _{\text{%
bottom}}=E_{\text{bottom}} $ with Eq.(\protect\ref{BandMinimum}) and $%
\protect\mu _{\text{cr}}$ is defined by Eq.(\protect\ref{Validity}).}
\label{FigGCon}
\end{figure}

The energy (\ref{Energy}) is minimized as%
\begin{equation}
E_{\text{bottom}}=-\frac{M\lambda ^{2}}{2\left( \hbar ^{2}-2\left\vert
J\right\vert M\right) }  \label{BandMinimum}
\end{equation}%
at the band minimum point%
\begin{equation}
k_{x}^{\text{bottom}}=\frac{M\lambda }{\hbar ^{2}-2\left\vert J\right\vert M}%
,\quad k_{y}^{\text{bottom}}=0,
\end{equation}%
as shown in Fig.\ref{FigBand}(a).

\begin{figure}[t]
\centerline{\includegraphics[width=0.48\textwidth]{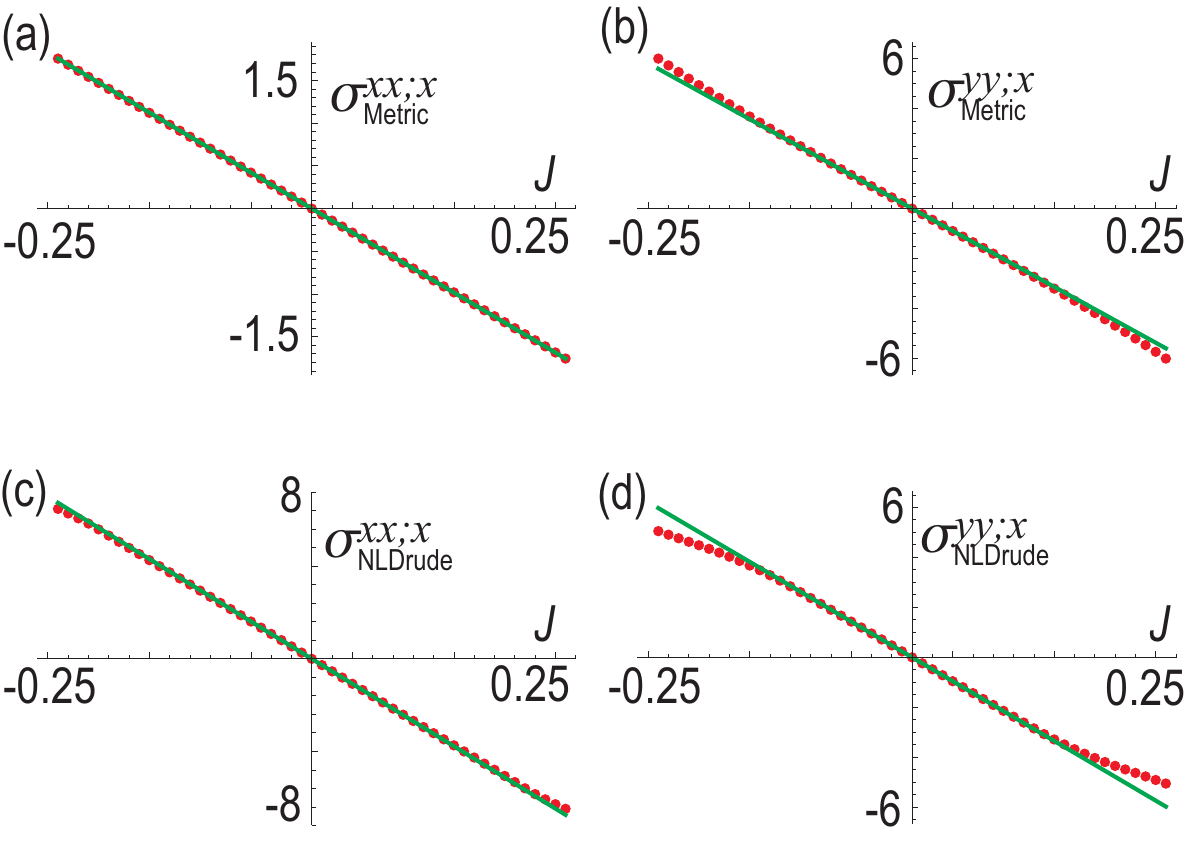}}
\caption{$J$ dependence of the nonlinear conductivity. (a) $\protect\sigma _{%
\text{Metric}}^{xx;x}$. (b) $\protect\sigma _{\text{Metric}}^{yy;x}$. (c) $%
\protect\sigma _{\text{NLDrude}}^{xx;x}$ and (d) $\protect\sigma _{\text{%
NLDrude}}^{yy;x}$. Their units are given in the caption of Fig.2. Red dots
indicate numerically obtained results, while green lines indicate
analytically obtained results. The horizontal axis is $J$ in units of $%
k_{0}^{2}$. We have set $\protect\mu =-0.2E_{0}$. See the definitions of $%
E_{0}$ and $k_{0}$ in the caption of Fig.1. }
\label{FigLinear}
\end{figure}

\textbf{Quantum-metric induced nonlinear conductivity:} We study the
longitudinal nonlinear conductivity induced by quantum metric. $\partial
G_{n}^{xx}/\partial k_{x}$ is antisymmetric with respect to $k_{x}$ for $J=0$%
. However, this antisymmetry is broken for $J\neq 0$. In addition, there is
no inversion symmetry in the energy (\ref{Energy}) for $J\neq 0$. As a
result, $\sigma _{\text{Metric}}^{xx;x}$ becomes finite for $J\neq 0$. $%
\sigma _{\text{Metric}}^{xx;x}$ is numerically calculated and is shown in
Fig.\ref{FigGCon}(a). The sign of the conductivity is reversed when the sign
of $J$ is reversed. It diverges at the Dirac point ($\mu =0$) as shown in
Fig.\ref{FigGCon}(a). $\sigma _{\text{Metric}}^{xx;x}$ is linear as a
function of $J$ as shown in Fig.\ref{FigLinear}(a).

We analytically calculate $\sigma _{\text{Metric}}^{xx;x}$ based on the
first-order perturbation theory in $J$, which is valid for $2M\left\vert
J\right\vert \ll \hbar ^{2}$. Actually, the conductivity is almost linear
for $2M\left\vert J\right\vert <\hbar ^{2}$ as shown in Fig.\ref{FigLinear}%
(a). The energy (\ref{Energy}) is expanded as%
\begin{equation}
E_{-}\left( \mathbf{k}\right) =-\lambda k+Jk^{2}\cos \theta \cos 2\theta +%
\frac{\hbar ^{2}k^{2}}{2M},
\end{equation}%
where we have introduced the polar coordinate of the momentum, $k_{x}=k\cos
\theta $ and $k_{y}=k\sin \theta $.

The Fermi surface is numerically obtained and shown in Fig.\ref{FigBand}(d).
In the vicinity of the Dirac point, there are two Fermi surfaces. In the
first-order of $J$, the Fermi surfaces at $E_{-}=\mu $ are determined by $%
k\left( \theta \right) =k_{\pm }\left( \theta \right) $ with%
\begin{equation}
k_{\pm }\left( \theta \right) =\frac{\lambda \pm \sqrt{\frac{2\mu \hbar ^{2}%
}{M}+\lambda ^{2}-2\mu J\left( \cos \theta +\cos 3\theta \right) }}{\frac{%
\hbar ^{2}}{M}-J\left( \cos \theta +\cos 3\theta \right) },  \label{kpm}
\end{equation}%
which is valid for%
\begin{equation}
\mu >-\frac{M\lambda ^{2}}{2\left( \hbar ^{2}+2M\left\vert J\right\vert
\right) }\equiv \mu _{\text{cr}}.  \label{Validity}
\end{equation}

The quantum-metric induced nonlinear conductivity is calculated as%
\begin{equation}
\sigma _{\text{Metric}}^{xx;x}=-\frac{e^{3}}{\hbar }\int d^{2}k\;f_{n}\frac{%
\partial G_{-}^{xx}}{\partial k_{x}}=\int_{k_{-}}^{k_{+}}kdk\int_{0}^{2\pi
}d\theta \frac{\partial G_{-}^{xx}}{\partial k_{x}}.
\end{equation}%
Up to the first order of $J$, we have%
\begin{equation}
\frac{\partial G_{-}^{xx}}{\partial k_{x}}=\frac{5\cos \theta \sin
^{2}\theta }{2\lambda k^{3}}+J\sin ^{2}\theta \frac{35\cos 4\theta -10\cos
2\theta -9}{8\lambda ^{2}k^{2}}.
\end{equation}%
By integrating it over $k$, we have%
\begin{align}
\sigma _{\text{Metric}}^{xx;x}& =-\frac{e^{3}}{\hbar }\int_{0}^{2\pi
}d\theta \frac{5\cos \theta \sin ^{2}\theta }{6\lambda }\left( \frac{1}{%
k_{-}^{2}}-\frac{1}{k_{+}^{2}}\right) \hspace{14mm}  \notag \\
& +J\sin ^{2}\theta \frac{35\cos 4\theta -10\cos 2\theta -9}{16\lambda
^{2}k^{3}}\left( \frac{1}{k_{-}}-\frac{1}{k_{+}}\right) .
\end{align}%
It is analytically calculated as%
\begin{equation}
\sigma _{\text{Metric}}^{xx;x}=-\frac{e^{3}}{\hbar }\frac{\pi J\sqrt{2\mu
\hbar ^{2}+M\lambda ^{2}}}{2\mu ^{2}\sqrt{M}\lambda }  \label{yyx1}
\end{equation}%
for $\mu <0$, and%
\begin{equation}
\sigma _{\text{Metric}}^{xx;x}=-\frac{e^{3}}{\hbar }\frac{\pi J}{2\mu
\lambda }\propto J  \label{yyx2}
\end{equation}%
for $\mu >0$. It is proportional to $J$. Hence, $J$ is measurable if the
magnitude of $\lambda $ is known. The formula well fits the numerical result
as shown in Fig.\ref{FigGCon}(a).

\textbf{Angle dependence:} Both $\sigma _{\text{Metric}}^{xx;x}$ and $\sigma
_{\text{NLDrude}}^{xx;x}$ are zero when the N\'{e}el vector is along the $x$
direction $\mathbf{n}=\left( 1,0,0\right) $ or the N\'{e}el vector is along
the $z$ direction $\mathbf{n}=\left( 0,0,1\right) $.

First, we study the angle dependence of the N\'{e}el vector along the $y$-$x$
plane by setting $\mathbf{n}=\left( \sin \Phi ,\cos \Phi ,0\right) $. In the
first-order of $J$, we analytically obtain $\sigma _{\text{Metric}%
}^{xx;x}(\Phi )=\sigma _{\text{Metric}}^{xx;x}(0)\cos \Phi $ and $\sigma _{%
\text{NLDrude}}^{xx;x}(\Phi )=\sigma _{\text{NLDrude}}^{xx;x}(0)\cos \Phi $.
Next, we study the angle dependence of the N\'{e}el vector along the $y$-$z$
plane by setting $\mathbf{n}=\left( 0,\cos \Theta ,\sin \Theta \right) $. In
the first-order of $J$, we analytically obtain $\sigma _{\text{Metric}%
}^{xx;x}(\Theta )=\sigma _{\text{Metric}}^{xx;x}(0)\cos \Theta $ and $\sigma
_{\text{NLDrude}}^{xx;x}(\Theta )=\sigma _{\text{NLDrude}}^{xx;x}(0)\cos
\Theta $.

\textbf{Nonlinear Hall conductivity:} Next, we calculate the quantum-metric
induced nonlinear Hall conductivity.

First, we have $\sigma _{\text{Metric}}^{xx;y}=0$, because 
\begin{equation}
2\frac{\partial G_{n}^{xx}}{\partial k_{y}}-\frac{\partial G_{n}^{xy}}{%
\partial k_{x}}\propto k_{y},
\end{equation}%
and its integration over $k_{y}$ vanishes.

On the other hand, we obtain a nontrivial result for $\sigma _{\text{Metric}%
}^{yy;x}$. The $\mu $ dependence of $\sigma _{\text{Metric}}^{yy;x}$ is
numerically obtained and shown in Fig.\ref{FigGCon}(b). It is linear as a
function of $J$ as in Fig.\ref{FigLinear}(b). It is analytically calculated
up to the first order in $J$ as%
\begin{equation}
\sigma _{\text{Metric}}^{yy;x}=-\frac{e^{3}\tau }{\hbar ^{2}}\frac{\pi
J\left( \mu \hbar ^{2}-5M\lambda ^{2}\right) }{4\mu \lambda ^{2}\sqrt{M}%
\sqrt{2\mu \hbar ^{2}+M\lambda ^{2}}}\propto J  \label{HallAna}
\end{equation}%
for $\mu <0$ and 
\begin{equation}
\sigma _{\text{Metric}}^{yy;x}=-\frac{e^{3}\tau }{\hbar ^{2}}\frac{5\pi J}{%
2\mu \lambda }\propto J
\end{equation}%
for $\mu >0$.

\textbf{Nonlinear longitudinal Drude conductivity:} The $\mu $ dependence of 
$\sigma _{\text{Drude}}^{xx;x}$ is shown in Fig.\ref{FigGCon}(c). It is
linear as a function of $J$ as shown in Fig.\ref{FigLinear}(c). It is
analytically calculated up to the first order in $J$ as%
\begin{equation}
\sigma _{\text{NLDrude}}^{xx;x}=-\frac{e^{3}\tau ^{2}}{\hbar ^{5}}6\pi J%
\sqrt{M}\sqrt{2\mu \hbar ^{2}+M\lambda ^{2}}\propto J
\end{equation}%
for $\mu <0$ and 
\begin{equation}
\sigma _{\text{NLDrude}}^{xx;x}=-\frac{e^{3}\tau ^{2}}{\hbar ^{5}}6\pi
JM\lambda \propto J
\end{equation}%
for $\mu >0$.

\textbf{Nonlinear transverse Drude conductivity:} The $\mu $ dependence of $%
\sigma _{\text{Drude}}^{yy;x}$ is shown in Fig.\ref{FigGCon}(d). It is
linear as a function of $J$ as in Fig.\ref{FigLinear}(d). It is analytically
calculated up to the first-order in $J$ as%
\begin{equation}
\sigma _{\text{NLDrude}}^{yy;x}=-\frac{e^{3}\tau ^{2}}{\hbar ^{5}}\frac{2\pi
J\sqrt{M}\left( 3\mu \hbar ^{2}+2M\lambda ^{2}\right) }{\sqrt{2\mu \hbar
^{2}+M\lambda ^{2}}}\propto J
\end{equation}%
for $\mu <0$%
\begin{equation}
\sigma _{\text{NLDrude}}^{yy;x}=-\frac{e^{3}\tau ^{2}}{\hbar ^{5}}4\pi
JM\lambda \propto J
\end{equation}%
for $\mu >0$.

\textbf{Discussions:} We have investigated the second-order nonlinear
conductivity of a system made of the $d$-wave altermagnet with the Rashba
interaction. We have focused on the problem to measure the direction of the N%
\'{e}el vector.

$\sigma _{\text{Metric}}^{xx;x}$, $\sigma _{\text{Metric}}^{yy;x}$, $\sigma
_{\text{NLDrude}}^{xx;x}$ and $\sigma _{\text{NLDrude}}^{yy;x}$ are
proportional to $J$ and have the same sign. Hence, $J$ is measurable
irrespective of $\tau $.

$\sigma _{\text{Metric}}^{xx;x}$\ and $\sigma _{\text{Metric}}^{yy;x}$\
diverge at the Dirac point $\mu =0$, while $\sigma _{\text{NLDrude}}^{xx;x}$%
\ and $\sigma _{\text{NLDrude}}^{yy;x}$\ are finite. Hence, $\sigma _{\text{%
Metric}}^{xx;x}$\ and $\sigma _{\text{Metric}}^{yy;x}$\ are dominant
comparing with $\sigma _{\text{NLDrude}}^{xx;x}$\ and $\sigma _{\text{NLDrude%
}}^{yy;x}$\ in the vicinity of the Dirac point.

In addition, the quantum-metric induced nonlinear conductivity $\sigma _{%
\text{Metric}}^{xx;x}$ and $\sigma _{\text{Metric}}^{yy;x}$\ become
significant in the dirty metal $\tau \rightarrow 0$. In this dirty regime, $%
\sigma _{\text{Metric}}^{xx;x}$ and $\sigma _{\text{Metric}}^{yy;x}$\ are
dominant comparing with $\sigma _{\text{NLDrude}}^{xx;x}$\ and $\sigma _{%
\text{NLDrude}}^{yy;x}$\ even away from the Dirac point.

The $x$-component of the N\'{e}el vector is measurable by observing $\sigma
_{\text{Metric}}^{yy;y}$. By combining the results of $\sigma _{\text{Metric}%
}^{xx;x}$ and $\sigma _{\text{Metric}}^{yy;y}$, the in-plane component of
the N\'{e}el vector is measurable.

The Berry-curvature dipole induced nonlinear Hall effect is another result
on the nonlinear conductivity. In the present system, the Berry curvature is
exactly zero except at the Dirac point because the system is gapless at the
Dirac point. Hence, there is no Berry-curvature dipole induced nonlinear
Hall effect in the present system.

This work is supported by CREST, JST (Grants No. JPMJCR20T2) and
Grants-in-Aid for Scientific Research from MEXT KAKENHI (Grant No. 23H00171).

\end{document}